# Unidirectional Synapse-Like Behavior of Zr/ZrO$_2$-NT/Au Layered Structure

Alexander S. Vokhmintsev
*NANOTECH Center*
*Ural Federal University*
Yekaterinburg, Russia
a.s.vokhmintsev@urfu.ru

Robert V. Kamalov
*NANOTECH Center*
*Ural Federal University*
Yekaterinburg, Russia
rvkamalov@urfu.ru

Anna V. Kozhevina
*NANOTECH Center*
*Ural Federal University*
Yekaterinburg, Russia
a.v.kozhevina@yandex.ru

Ilya A. Petrenyov
*NANOTECH Center*
*Ural Federal University*
Yekaterinburg, Russia
ilyapetrenyov@mail.ru

Nikolay A. Martemyanov
*NANOTECH Center*
*Ural Federal University*
Yekaterinburg, Russia
n.a.martemianov@urfu.ru

Ilya A. Weinstein
*NANOTECH Center*
*Ural Federal University*
Yekaterinburg, Russia
i.a.weinstein@urfu.ru

*Abstract* — Zirconia nanotubular layer with an outer tube diameter ≈ 25 nm was synthesized by potentiostatic anodization. The Zr/ZrO$_2$-NT/Au memristive structure is fabricated using stencil mask and magnetron sputtering techniques. Current-voltage characteristics are measured in full cycles of resistive switching with varying parameters of the applied harmonic voltage. An equivalent circuit with unidirectional electrical conductivity for the studied structure is proposed. Estimates of the electrical resistance of memristors in high- and intermediate resistivity states are performed. The high synaptic plasticity of memristors based on the Zr/ZrO$_2$-NT/Au structure is shown.

## I. Introduction

Developing of hybrid schemes that have analogous-digital architecture combined with crossbar structure is one of the promising directions for fabrication of neural chips, neurocomputing systems and artificial neural networks (ANNs) nowadays [1, 2]. These schemes have neurons based on common integral CMOS transistors. There, metallic conducting structures are axons and dendrites, wherein, two-terminal memristive elements of commutation matrix serve as synapses. Memristors are formed at crossing points of conductors which connect pre- and postsynaptic neurons.

Biological systems have synaptic weight of connection between neurons regulated by ionic flow. It is many known values of synaptic weights and possibility to change them that enables learning and functioning in biological systems [3, 4]. It can be noted that memristor is a two-terminal electron device with synaptic plasticity or ability to change synaptic weight, since its conductivity depends on total charge flowing [2].

As a rule, the connections between neurons in biological systems and, consequently, in the existing variety of ANN architectures are unidirectional in terms of excitation propagation [6]. Thus, fabrication and studying of memristive structures of metal-insulator(semiconductor)-metal (MIM structures) with unilateral electrical conductivity appear to be promising directions.

It is known that memristive behavior is caused by mobility of cation and anion vacancies, etc. in insulator layer. Its thickness and presence of defects determine electrical resistance in low- (LRS), high- (HRS) and intermediate resistivity states (IRS) [1, 3]. There is an effect of reversible resistance switching for several MIM structures based on transition metals (TiO$_2$, ZrO$_2$, HfO$_2$, ets.) obtained by anodizing or electrochemical oxidation [2, 7–10].

Therefore, the aim of this work was to synthesize, characterize and study static current-voltage (CV) characteristics of layered structure with unidirectional electrical conductivity based on anodic zirconia.

## II. Experimental section

### A. Fabrication of Zr/ZrO2-NT/Au structure

The Zr/ZrO$_2$-NT/Au memristor structure was fabricated by anodizing of Zr metal and magnetron sputtering of Au top layer on the surface of the oxide layer.

The Zr metal foil of 120 μm thick was used to synthesize the ZrO$_2$-NT layer by anodization. The Zr foil was pre-washed with acetone, etched in a solution of HF:HNO$_3$:H$_2$O = 1:6:20 acids, intensively washed with distilled water and dried in air. Synthesis of the oxide layer was carried out in a two-electrode cell at a constant voltage of 20 V for 15 min, where Zr foil was an anode and a steel plate as a cathode. The electrolyte was an ethylene glycol solution containing 5 wt. % H$_2$O and 1 wt.% NH$_4$F. All chemical reagents were analytical grade purity.

Gold contacts were sputtered using a stencil mask of 100 holes with a diameter of 1 mm by magnetron unit Q150T ES Quorum Technologies on the surface of the obtained Zr/ZrO$_2$-NT sample. Deposition was carried out for 500 s, as a result of which the thickness of the Au contact was 50 nm. Thus, 100 samples of memristors of Zr/ZrO$_2$-NT/Au sandwich-structure were formed.

### B. Characterization of Zr/ZrO2-NT/Au structure

The morphology of the synthesized structures was study with a scanning electron microscope SIGMA VP Carl Zeiss.

The current-voltage measurements in the static operation mode were carried out with the original installation [11]. The experimental setup includes a Cascade Microtech MPS150 micro-probing station and a modular device Source Measure Unit (SMU) PXI-4130 from National Instruments. The Zr foil was grounded, and the harmonic signal $U(t) = U_0 + U_m \sin(2\pi f t)$ with frequency of $f = 0.01$ Hz was applied to the Au contacts. The amplitude of the applied voltage $U_m$ and the bias voltage $U_0$ were varied to change $U(t)$ in the range from –8 V to 9 V. Automation of the procedure for measuring and storing experimental data was carried out using VI "CVC" virtual instrument in the graphical design platform LabVIEW [11].

## III. RESULT AND DISSCUTIONS

### A. Characterization of ZrO2-NT layer

Fig. 1 shows a typical SEM image of the surface of a synthesized zirconia layer (a) and its cross-section (b). As a result of electrochemical oxidation of zirconium foil a self-ordered array of zirconia nanotubes with an external diameter of ≈ 25 nm and a length of ≈ 3 μm was formed. In our earlier study was shown that as-grown layer of $ZrO_2$-NT contains 90% of tetragonal and 10% of monoclinic phases [12].

### B. Current-voltage measurements of Zr/ZrO2-NT/Au structure

Fig. 2 shows the time dependence of the current through the memristor and the applied harmonic voltage $U(t)$ ($U_0 = 2$ V, $U_m = 6$ V) for 4 measurement periods. It is seen when $U(t) < 0$ V the reverse current through the memristor is $I_S < 0.05$ mA. The current jumps in the region of the maximum values of the applied voltage are observed for $U(t) > 0$ V. This behavior is typical for electronic devices with unidirectional conductivity, i.e. diodes. It is known that the work function of metal contacts is $W_{Au} = 5.1$ eV, $W_{Zr} = 4.05$ eV [13] and $W = 4.6$ eV for stoichiometric $ZrO_2$ [14]. Comparing the $W$ values for the materials of the investigated MIM structure, we can conclude that a Schottky barrier is formed at the Au/$ZrO_2$-NT interface, and the Zr/$ZrO_2$-NT contact is ohmic.

From Fig. 2 it is also seen that the maximum value of $I$ increased monotonically in the range of 73 – 150 mA for each subsequent period of the $U$ change. This fact indicates about a 2-fold decrease of the electrical resistance of the studied structure under the given experimental conditions. The Fig. 3a shows the CV curves corresponding to the experimental data presented in Fig. 2. It is seen that for $U > 0$ V a typical CV loops similar to MIM structures with bipolar resistive switching are registered [1, 2]. Otherwise, for $U < 0$ V a typical branch of CV curve representative of a diode under reverse voltage is recorded. In the inset to Fig. 3a, the connection scheme for investigated MIM structure to the measuring equipment is shown.

Basing on the above mentioned the inset on Fig. 3b shows the equivalent circuit for the studied Zr/$ZrO_2$-NT/Au sandwich-structure. The electrical circuit consists of a series connection of a *VD* Schottky diode and a *M* memristor.

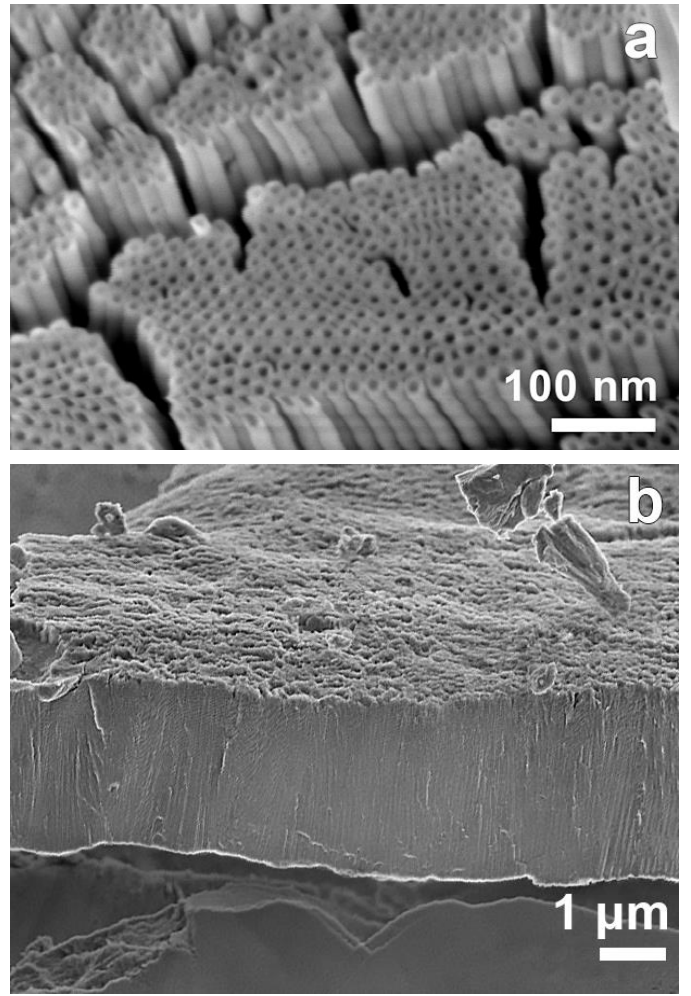

Fig. 1. Zirconium dioxide nanotubular structure: (a) top view; (b) cross-sectional view.

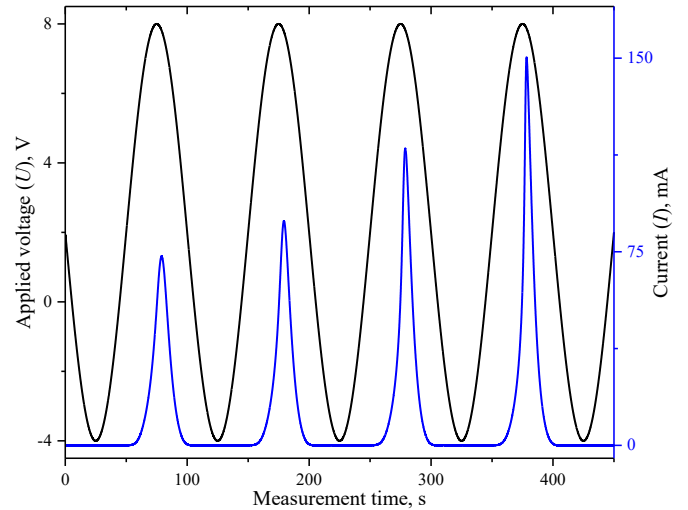

Fig. 2. The dependencies of the flowing current and the applied voltage on time for the Zr/$ZrO_2$-NT/Au structure.

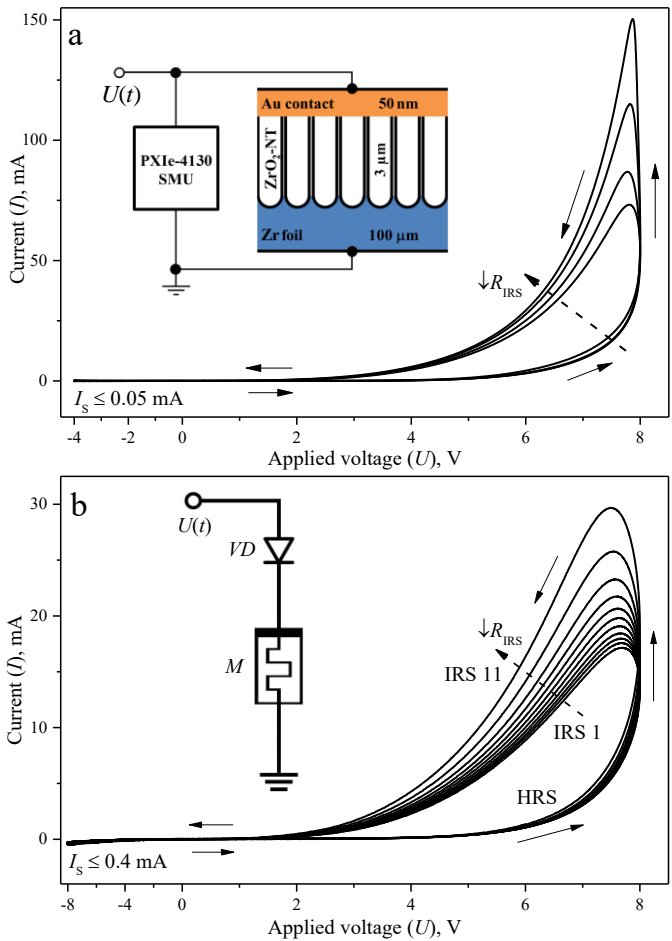

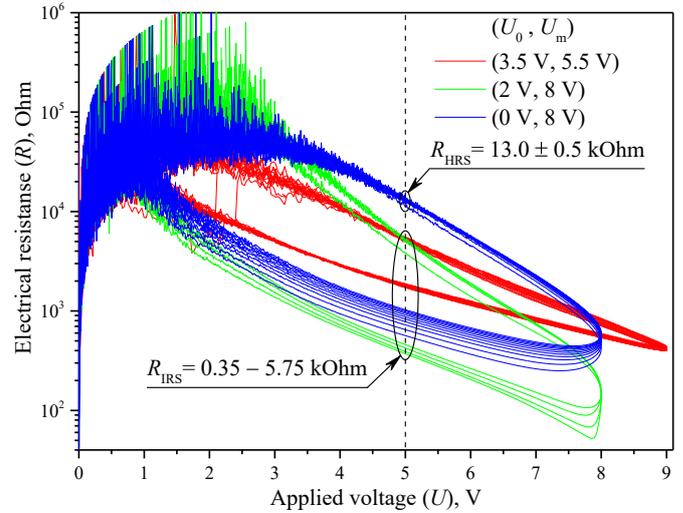

Fig. 3. B. Current-voltage characteristics of full resistance switching cycles of Zr/ZrO$_2$-NT/Au structure under $U(t) = U_0 + U_m \sin(2\pi f t)$ harmonic influence for: (a) $U_0 = 2$ V, $U_m = 6$ V and (b) $U_0 = 0$ V, $U_m = 8$ V. Arrows show direction of registration. Insets shows: (a) wiring diagram for studied MIM structure and PXI instrument module; (b) its equivalent electric circuit.

So, it can be stated that electrical conductivity of oxide layer is changed under harmonic $U(t)$ voltage applied to studied structure due to movement of intrinsic defects relative to their original location. Pointed out defects positions correspond to several IRSs in memristors (see Fig. 3a) and responsible for their synaptic weight. These researches showed that maximum backward voltage of studied MIM structure was $U_{BR} > 8$ V. Thus, a series of measurements for memristors under $U_m = 8$ V were carried out in order to transfer them to nominal HRS (IRS with the highest resistance for certain experimental conditions) with subsequent switching to IRS. Fig. 3b shows eleven full resistivity switching cycles for the conditions. It can be derived from CV curves that backward voltage decrease to $U = -8$ V causes reverse current rise to $I_S \leq 0.4$ mA. A reproduction of CV characteristic branches in HRS and monotone increase of straight current through structure in IRSs for every consecutive measurement cycle are registered for $U > 0$ V.

Fig. 4 shows dependencies of electrical resistance $R$ versus $U(t)$ voltage applied to Zr/ZrO$_2$-NT/Au structure for straight branches of CV characteristics measured for memristors in different intermediate state. It is important to note that considerable $R$ values fluctuations are presented for $U < 5$ V. The fact can be attributed to current registration accuracy for certain experimental conditions. Therefore, $R$ evaluation in HRS and IRS were conducted under $U = 5$ V. Values of $R_{HRS} = 13.0 \pm 0.5$ kOhm and $R_{IRS}$ from $\approx 5.75$ to $\approx 0.35$ kOhm were calculated according to CV characteristic of studied structure. It was substantiated that there are significant range of synaptic weights variation and high synaptic plasticity in studied samples due to $R_{HRS}$ / $R_{IRS}$ resistance ratio varying from $\approx 2$ to $\approx 37$.

Fig. 4. Plot of electrical resistance versus applied voltage for Zr/ZrO$_2$-NT/Au structure. $U(t)$ parameters are shown in the legend.

## IV. CONCLUSIONS

ZrO$_2$-NT nanotubular layer was synthesized by zirconium foil anodizing in 5 wt. % H$_2$O and 1 wt. % NH$_4$F ethylene glycol solution under constant 20 V voltage for 15 min. Scanning electron microscopy analysis showed that oxide layer had $\approx 3$ μm thickness and consisted of nanotubes with $\approx 25$ nm diameter.

Zr/ZrO$_2$-NT/Au memristive structures with 1 mm diameter gold contacts were fabricated by magnetron sputtering technique using the stencil mask. CV characteristics of full resistance switching cycles under varied applied harmonic voltage parameters were studied. Equivalent electrical circuit with unidirectional conductivity was proposed for MIM structure consisting of Schottky diode and memristor series connection according to experimental dependencies analysis. Values of $R_{HRS} = 13.0 \pm 0.5$ kOhm and $R_{IRS} = 5.75 - 0.35$ kOhm electrical resistivity were calculated for memristors in HRS and IRS. It was found that investigated structures had high synaptic plasticity according to $R_{HRS}/R_{IRS} = 2 \div 37$ ratio range. The study showed prospects for use of Zr/ZrO$_2$-NT/Au based memristor with unidirectional electrical conductivity for solid state modeling of synapses in artificial neural systems.

ACKNOWLEDGMENTS

The work was supported by Act 211 Government of the Russian Federation, contract № 02.A03.21.0006. Contribution to the study from R.V.K. was funded by the RFBR according to the research project № 18-33-01072. A.S.V. and I.A.W. thank Minobrnauki initiative project № 16.5186.2017/8.9 for support.

## References

[1] S. H. Jo, T. Chang, I. Ebong, B. B. Bhadviya, P. Mazumder, and W. Lu, "Nanoscale memristor device as synapse in neuromorphic systems," Nano Lett., vol. 10, pp. 1297-1301, April 2010.

[2] J. J. Yang, D. B. Strukov, D. R. Stewart, "Memristive devices for computing," Nat. Nanotechnol., vol. 8, pp. 13-24, January 2013.

[3] G. Indiveri, B. Linares-Barranco, R. Legenstein, G. Deligeorgis, and T. Prodromakis, "Integration of nanoscale memristor synapses in neuromorphic computing architectures," Nanotechnol., vol. 24, pp. 27-98, September 2013.

[4] G. Indiveri, E. Chicca, and R. Douglas, "A VLSI array of low-power spiking neurons and bistable synapses with spike-timing dependent plasticity," IEEE Trans. Neur. Netw., vol. 17, pp. 211-221, February 2006.

[5] S. Song, K.D. Miller, and L.F. Abbott, "Competitive Hebbian learning through spike-timing-dependent synaptic plasticity," Nat. Neurosci., vol. 3, pp. 919-926, March 2000.

[6] G. A. Carpenter, S. Grossberg, N. Markuzon, J. H. Reynolds, and D. B. Rosen, "Fuzzy ARTMAP: A Neural Network Architecture for Incremental Supervised Learning of Analog Multidimensional Maps," IEEE Trans. Neur. Netw., vol. 3, pp. 698-713, September 1992.

[7] T. V. Kundozerova, "Unipolyarnoe rezistivnoe perekluchenie v structurakh na osnove oksidov niobiya, tantala i zirconiya," Kand, Diss. [Unipolar resistive switching in structures based on the oxides of niobium, tantalum and zirconium, PhD] Petrozavodsk, 2013.

[8] K. Miller, K. S. Nalwa, and A. Bergerud, "Memristive Behavior in Thin Anodic Titania," IEEE Electron. Dev. Lett., vol. 31, p. 737, May 2010.

[9] J. Yoo, K. Lee, and A. Tighineanu, "Highly ordered TiO2 nanotube-stumps with memristive response," Electrochem. Comm., vol. 34, pp. 177-180, June 2013.

[10] A. S. Vokhmintsev, I. A. Weinstein, R. V. Kamalov, and I. B. Dorosheva, "Memristive Effect in a Nanotubular Layer of Anodized Titanium Dioxide," Bull. Rus. Acad. Sci., Phys. Russia, vol. 78, pp. 1176–1179, January 2014.

[11] A. O. Gryaznov, I. B. Dorosheva, A. S. Vokhmintsev, R. V. Kamalov, and I. A. Weinstein, "Automatized complex for measuring the electrical properties of MIM structures," Int. Siberian Conf. Control Comm., pp. 749-772, June 2016.

[12] A. V. Kozhevina, A. S. Vokhmintsev, R. V. Kamalov, N. A. Martemyanov, A. V. Chukin, and I. A. Weinstein, "Optical absorption edge parameters of zirconium dioxide nanotubular structures," IOP Conf. Series: J. Phys., p. 917, November 2017.

[13] D.E. Eastman, "Photoelectric work functions of transition, rare-earth, and noble metals," Phys. Rev. B: Condens. Matter, vol. 2, pp. 132-143, April 1970.

[14] A. A. Knizhnik, I. M. Iskandarova, and A. A. Bagatur'yants, "Impact of oxygen on the work functions of Mo in vacuum and on ZrO2," J. Appl. Phys., vol. 97, December 2005.